\documentclass[preprint]{aastex}   	
\usepackage{geometry}                		
\usepackage{graphicx}				
\usepackage{natbib}	

\title{Coded Aperture Imaging in High-Energy Astrophysics}
\author{Jo\~ao Braga}
\affil{National Institute of Space Research -- INPE, Avenida dos Astronautas 1758, \\ S\~ao Jos\'e dos Campos, SP, Brazil, 12227--010}
\email{joao.braga@inpe.br}

\shorttitle{Coded Aperture Imaging}
\shortauthors{J. Braga}

\date{}							

\slugcomment{accepted by PASP as an invited review article}

\newcommand{\ts}{\thinspace}

\begin{abstract}
Hard X-ray and low-energy gamma-ray coded-aperture imaging instruments have been highly successful as high-energy surveyors and transient-source discoverers and trackers over the past decades. Albeit having relatively low sensitivity as compared to focussing instruments, coded-aperture telescopes still represent a very good choice for simultaneous, high cadence spectral measurements of individual point sources in large source fields. Here I present a review of the fundamentals of coded-aperture imaging instruments in high-energy astrophysics. Emphasis is on fundamental aspects of the technique, coded-mask instrument characteristics, and properties of the reconstructed images. 
\end{abstract}


\begin{document}
\maketitle

\section{Introduction}
\label{sec:intro}
 
In usual coded-aperture imaging (hereinafter CAI) instruments, a plate (referred to as a {\it mask\/}) built with a pattern of elements that are either (semi)opaque or transparent to the radiation of interest is placed at a certain distance from, and usually parallel to, a position-sensitive detector plane system (hereinafter PSDP). In essence, the mask spatially encodes the incoming radiation in a unique way for each direction in the field-of-view (FOV) of the instrument, so that a suitable processing of the radiation intensity (usually photon counts) distributed over the PSDP for a given integration time is able to reproduce the intensity in each direction, which provides an image of the FOV for a certain energy range.

CAI has been a very successful technique for measuring the angular locations and the electromagnetic fluxes and spectra of astronomical objects above energies for which geometric optics standard technology is not able to effectively achieve an ordered reflection of photons. Albeit being inherently a low signal-to-noise imaging technique due to its non-focussing nature, CAI is still a widely used method for carrying out X-ray and low energy gamma-ray monitoring and surveying of astrophysical sources due essentially to two general shortcomings of grazing incidence (Wolter) telescopes: (a) the highest energies achievable are still below 100 keV and (b) the fields of view are typically not much greater than about 20 arcmin in diameter. Coded mask instruments, on the other hand, allow imaging over very wide fields (which can be a significant fraction of the sky) and up to energies of several hundred keV or even a few MeV. The technology to develop focussing telescopes at energies above $\sim$15\ts keV has only recently been mastered and implemented, especially in the NuSTAR observatory \citep{2013ApJ...770..103H}. While providing much higher sensitivities and better angular resolution in general, focussing instruments at high X-ray energies require very long focal lengths and sophisticated control and alignment systems, whereas CAI instruments can be compact and relatively easy to build, making them good options for wide-field hard X-ray and low energy $\gamma$-ray monitors of the highly variable and transient source populations, still largely unidentified.

Coded mask satellite instruments that have made significant contributions to astronomy include SL2/XRT on {\it Spacelab 2\/} \citep{1987JBIS...40..159E}, SIGMA/{\it GRANAT\/} \citep{1991AdSpR..11..289P}, the WFCs on {\it BeppoSAX\/} \citep{1997A&AS..125..557J}, WFM/{\it HETE\/} \citep{2003AIPC..662....3R}, instruments on the {\it INTEGRAL\/} satellite \citep{1995ExA.....6...71W}, and BAT/{\it Swift\/} \citep{2004NewAR..48..431G}.

In this article, I review the fundamentals of two-dimensional CAI for astronomical applications, in which the objects are considered to be at infinity. CAI is also used in other applications such as medical imaging \citep{2006NIMAS.563..146P, 2007NIMAS.573..122P} and nuclear material detection for national security \citep{2012IEEE..30P}, in which the various distances from the objects to the instrument have to be taken into account in the reconstruction algorithms. These further implementations of CAI are not treated here.

After the first CAI space experiments were developed and flown, comprehensive reviews of astronomical spatial multiplexing imaging techniques, including CAI, were published  \citep{1987SSRv...45..349C, 1995ExA.....6....1S}. In the Web, an excellent description of CAI  is available at \citet{1992NASACAIintzand} and references therein. The focus of this article is to bring the field up to date and review some aspects of the technique that were not specifically covered before, such as some special features of coded-mask instruments and some statistical properties of the reconstructed images. I also comment on important results obtained by more recent coded-mask space instruments. 

\section{The concept}
\label{sec:concept}

The coded aperture concept was introduced independently by \citet{1968ApJ...153L.101D} and \citet{1968PASAu...1..172A} as an extension of a simple pinhole camera. In his paper, Ables prophetically stated that, with an X-ray multiple-pinhole camera, ``broad surveys with sufficient resolution to virtually assure positive optical identification of any source detected are possible". This came to be dramatically true with missions such as {\it GRANAT, BeppoSAX, INTEGRAL\/} and {\it Swift}, and future survey missions using CAI can still make important contributions.

\begin{figure}[!ht]
\centering
\includegraphics[width=0.9\hsize,angle=0]{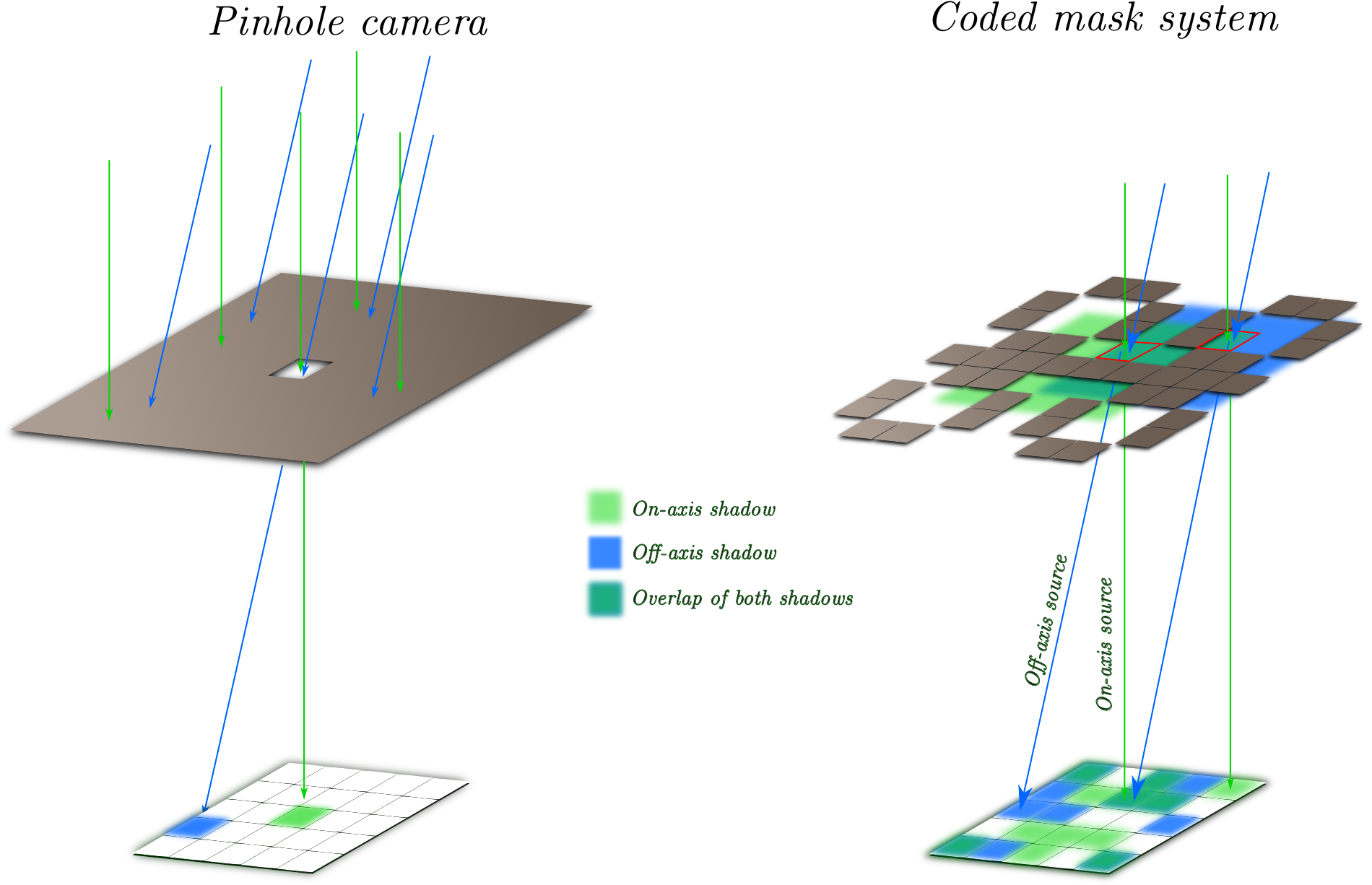}
\caption{{\it Left:\/} a pinhole camera or a single-open-cell coded-mask imager. The detector plane registers an image of the source field with relatively low significance peaks, since the detecting area corresponds to only one mask element. {\it Right:\/} a coded-mask imager. The detector plane registers an overlapping set of multiple images (or shifted shadowgrams of the mask pattern), each set associated with one point source. In this case the actual image has to be produced by a suitable algorithm which is essentially a cross-correlation of the combined shadowgram with the mask pattern. The peaks in the reconstructed image (not shown here) have much higher significance, since the number of detected source photons is multiplied by the number of open cells of the basic mask pattern; see section 6.}
\label{fig:mask}
\end{figure}

The basic idea of an astronomical coded-mask imager is to mount a plate with a pattern of (semi)opaque and transparent elements or ``cells" (the {\it mask\/}) parallel to a PSDP so that the mask covers the aperture through which the incoming radiation reaches the detectors. An observation of a point source multiplies the illuminated area of the PSDP, with respect to an opaque mask with a single transparent cell, by the number of open cells projected in that particular direction. This clearly preserves the geometric angular resolution of the single cell case, given by the cell size divided by the distance from the mask to the PSDP, while in general greatly enhances the sensitivity of the imager, since the number of source photons detected is proportional to the detecting area illuminated by the source. In high-energy astrophysics, this is of utmost importance due to the fact that most observations are subjected to high background radiation. As pointed out by \citet{2003astro.ph..2354S}, the point source response function of a CAI instrument usually extends over the full area of the detector plane, and the multiple images of the source actually become an inverted shadowgram of the opaque elements of the mask (see Figure \ref{fig:mask}).

If there are several sources in the FOV and a spatial resolution $\Delta s \ll m$ in the detector plane (where $m$ is the mask cell linear size), what gets imprinted in the PSDP, in addition to the background distribution, is a set of overlapping shadows of the mask, each one shifted according to the position of the source in the FOV. This bears almost no resemblance with an actual image of the sky, but all the information required to measure the intensities and positions of the sources is contained in the PSDP intensity map, provided the mask pattern is known. Therefore, an essential feature of CAI is that it is a two-step technique: (i) the acquisition of an event distribution (illumination pattern or shadowgram) over the PSDP, (ii) the application of a post-processing technique to produce, or ``reconstruct", the actual image. This works even with large open areas in the mask (shaped by adjacent open cells), since what actually matters is the ability to define the edges of the illuminated regions in the PSDP.

\section{The detector plane spatial resolution}

Any X-ray and/or low-energy $\gamma$-ray PSDP has a finite spatial resolution given by the mechanical and electronic solutions employed to determine the photon interaction locations within the detector material. In recent instruments, solid state composite semiconductor materials such as CdZnTe or CdTe have been used as X and low-energy $\gamma$-ray detectors due to their high photoelectric absorption and good energy resolution at room temperatures. To make up a PSDP, one has to use either an array of individual detectors or detectors with a readout system that provides determination of the two-dimentional position of interaction in the detecting material with a certain accuracy. One can also employ a combination of the two, with position-sensitive detectors tiled to form the PSDP.

The main consequence of having a less-than-perfect PSDP is that the sizes of the individual openings in the mask cannot be as small as desired to achieve the best possible angular resolution, because if the mask elements are significant smaller than the spatial resolution $\Delta s$, other crucial parameters such as instrument sensitivity and source position accuracy are severely degraded due to the blurring of the edges of the mask element shadows in the PSDP. For high-energy radiation, i.e.\ X and $\gamma$ rays, it is noteworthy that diffraction is negligible even with very small pinholes, provided that $\lambda \ll m$, where $m$ is the typical pinhole linear size.

\citet{2008ApOpt..47.2739S} has confirmed through simulations that a good approximation for the actual angular resolution $\Delta\theta_{ar}$ of a coded-mask telescope is  given by
\begin{equation}
\Delta\theta_{ar}^2 = (m/l)^2 + (\Delta s/l)^2,
\end{equation}
where $l$ is the mask-PSDP separation (here we assume that we use the same definition for all resolutions, e.g.\ the full width at half maximum -- FWHM -- of a distribution). Therefore, if $m \ll\Delta s, \Delta\theta$ gets dominated by $\Delta s$ for a given $l$, so that using a mask with very small openings, besides being potentially difficult to build, is of little effect in the resolution. In addition, the source location accuracy $\Delta\theta_{sla}$ can be estimated \citep{2008ApOpt..47.2739S} by dividing $\Delta\theta_{ar}$ by a factor proportional to the signal-to-noise ratio $S\!N\!R$, or statistical significance, of the point source detection (to be discussed in section \ref{sec:snr}):
\begin{equation}
\Delta\theta_{sla} \propto \frac{\Delta\theta_{ar}}{S\!N\!R}.
\end{equation}
As the $S\!N\!R$ strongly depends on the $\Delta s/m$ ratio due to the blurring effect, $\Delta\theta_{sla}$ degrades significantly when $m \ll \Delta s$. \citet{2008ApOpt..47.2739S} also shows that, under reasonable assumptions about count statistics and PSDP spatial uniformity, the lowest source angular position uncertainty is obtained for $m \approx \Delta s$, which is in good agreement with  empirical results and simulations. For this reason, most masks in CAI instruments are built with basic cells that approximately match the spatial resolution on the PSDP or are larger by a factor of a few ($m \gtrapprox \Delta s)$.

\section{Geometry and fields of view}

It is always desirable that astronomical CAI instruments have mask areas at least equal to the PSDP areas, so that a point source in the direction of the instrument axis will cast a shadow of the mask over the entire PSDP, maximizing the source-detecting area. In other incidence directions, the total illuminated area on the PSDP will depend on the geometry of the instrument. It is important to distinguish between the {\it fully-coded field of view\/} (FCFOV), within which all the photons from the incoming directions that would reach the PSDP pass through the mask and, in so doing, cast a mask shadow over the entire PSDP, from the {\it partially-coded field of view\/} (PCFOV), for which a fraction of these photons will either be blocked by side-shielding material or reach the PSDP without going through the mask (will pass beyond the edges of the mask).

\begin{figure}[!ht]
\centering
\includegraphics[width=0.55\hsize,height=5.5cm,angle=0]{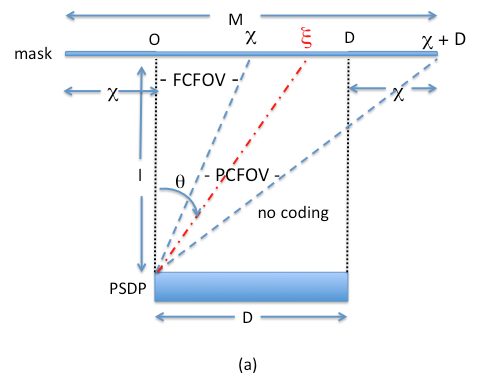}
\hspace{0.2cm}
\includegraphics[width=0.75\hsize,height=5.5cm,angle=0]{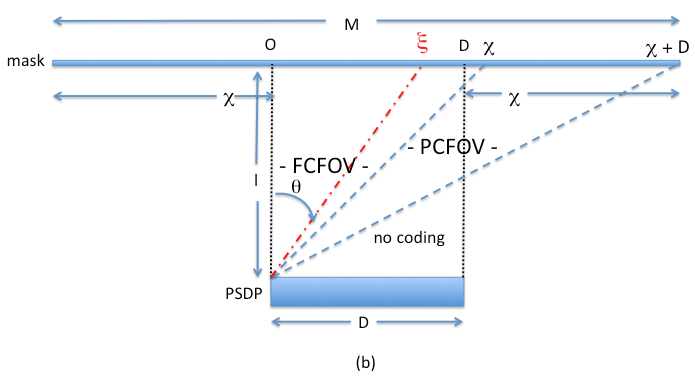}
\caption{a cutaway view (in a plane parallel to the instruments's axis) of a CAI instrument showing one half of the fields of view in the azimuthal direction associated with this cut (for an instrument with a cylindrical symmetry, this reflects all central planes). In (a) the dimension of the mask $M$ in the direction shown is less than twice the PSDP dimension, whereas in (b) $M \ge 2D$. If $M = D$, the FCFOV is zero. $\xi$ corresponds to a generic direction (see text) that is in the PCFOV in (a) but inside the FCFOV in (b). For clarity, no shieldings or structural parts are depicted; the dotted lines do not correspond to any structure.}
\label{fig:fov}
\end{figure}

In Figure \ref{fig:fov} we show the general picture for two different geometries: in (a) the mask linear dimension $M$ is less than twice the linear dimension of the detector plane $D$ in the azimuthal direction of this cutaway view, whereas in (b), $M \geq 2D$. For a given cutaway ``vertical'' plane of the instrument, we can define the relevant ranges of incidence directions by marking fiducial points in the mask cut. A given direction $\theta$ of incoming radiation, corresponding to point $\xi$, is clearly $\tan^{-1} (\xi/l)$. Half of the FCFOV in this direction (the other half is completely symmetrical to the left in the figure) clearly spans $0$ to $\chi$, whereas half of the PCFOV spans $\chi$ to $\chi+D$. In (a), direction $\theta$ is within the PCFOV, since $\xi$ is beyond point $\chi$ and so there will be no full coding of the incoming radiation by the mask. In (b), however, direction $\theta$ is inside the FCFOV since, in this case, $\chi > \xi$. Therefore, if one uses a mask basic pattern with linear size $D$ that does not repeat itself in any permutation of its cells, the difference between cases (a) and (b) is that in (a) there will be no direction ambiguities in the FCFOV, since all directions will cast different shadowgrams onto the detector plane ($\chi < D/2$), whereas in (b) such ambiguities will happen for directions corresponding to $D/2 \le \xi < \chi$.

\begin{figure}[!ht]
\centering
\includegraphics[width=0.65\hsize,height=5.5cm,angle=0]{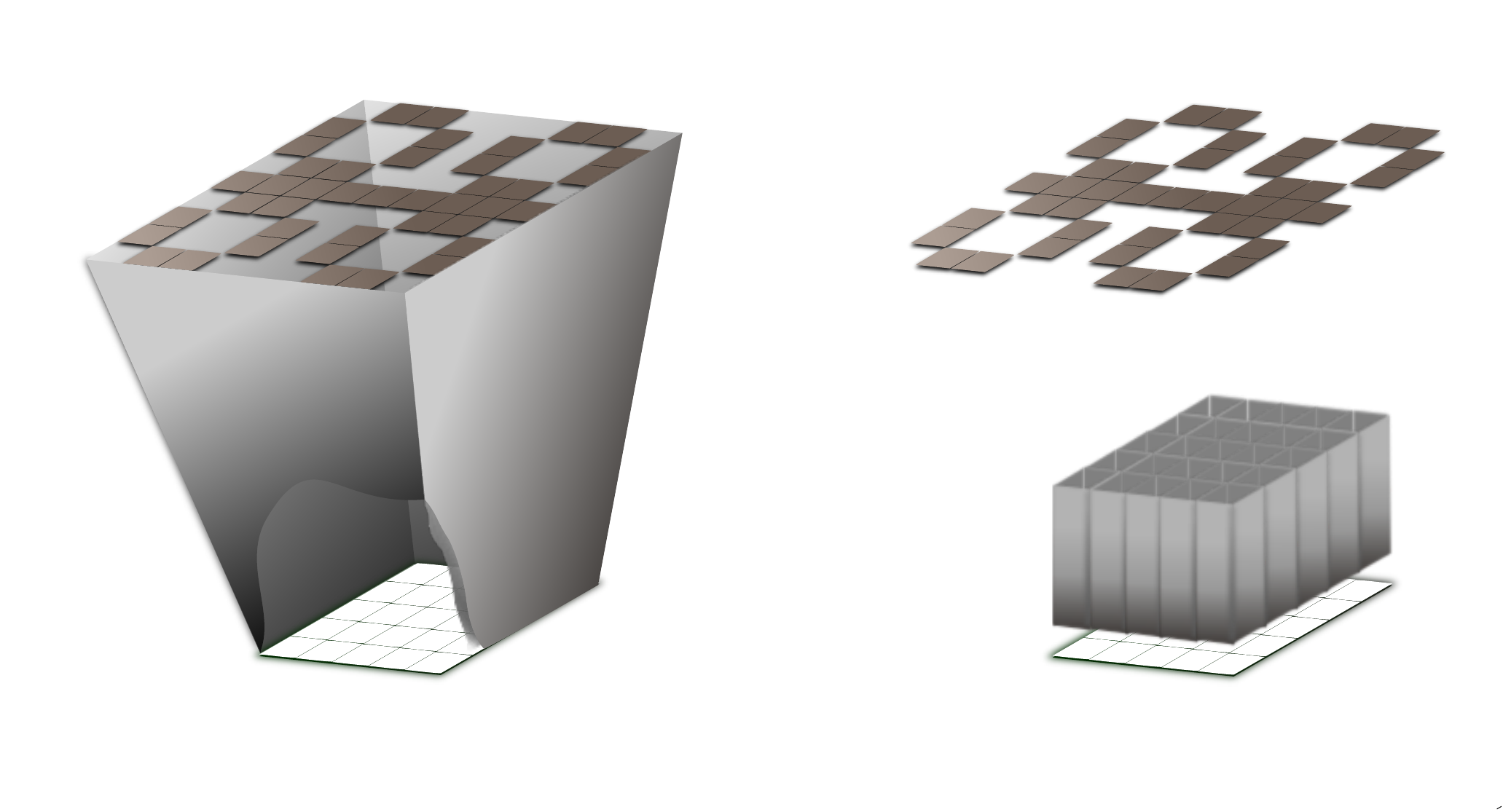}
\caption{Two different usual implementations of a coded-mask camera. {\it Left\/}: walls of shielding material are placed so as to join the sides of the detector plane and the mask, avoiding photons that do not pass through the mask to reach the detectors. {\it Right\/}: a collimator is placed in the front of the detector plane to define the FOV. The height of the collimator walls are such that the mask covers exactly the entire FOV. Structural parts are not shown for clarity.}
\label{fig:imp}
\end{figure}

In order to define the FOV of a CAI instrument, several approaches can be used. The most common one is to surround the back and sides of the PSDP with shielding materials end extend those all the way to the mask edges, as in a conventional camera. This is depicted in the left side of Figure \ref{fig:imp}. Several space experiments have used this approach. A major advantage is that there will be a range of incidence directions ($\xi < \chi$ in Figure \ref{fig:fov})  for which the FCFOV will have maximum sensitivity, since the whole detection area is being illuminated. The disadvantage here is that part of the FOV will be partially coded, since in a range of directions ($ \chi < \xi < \chi + D$ in Figure \ref{fig:fov}) the mask shadowgrams cast over the PSDP will be incomplete. One extreme case of this configuration, e.g.\ the WFCs on {\it BeppoSAX\/} \citep{1997A&AS..125..557J}, is when the mask and the PSDP have equal sizes, so the entire FOV is partially coded. Experiments with relatively large PCFOVs can in principle have problems with the image reconstruction process, especially in cases in which one or more bright sources happen to fall in one of these positions in the sky. However, experiments of this type can obtain extremely significant results, as was the case with the above-mentioned WFCs on {\it BeppoSAX\/} \citep{2007A&A...472..705V}.

Another approach, such as the one used in the EXITE \citep{1989ITNS...36..871B} and protoMIRAX \citep{2015A&A...580A.108B, 2016MNRAS.459.3917C} experiments, is to avoid the use of side shieldings between mask and detector altogether and implement instead a grid of shielding blades (a collimator) in front of the PSDP to define the FOV in a way that the spacing between the blades corresponds to the detector spatial resolution (Figure \ref{fig:imp}, right). This idea was first discussed by \citet{1979MNRAS.187..633P}, who introduced an ``egg-crate'' collimator in their coded-mask telescope in the {\it Spacelab-2\/} mission. One particularly interesting use of this approach happens when the PSDP is made of individual detectors, each one with no position sensitivity. In these cases, as the incidence angle increases with respect to the axial direction, the blades will cast equal shadows in all resolution elements (the detectors themselves) in the PSDP, so the coding by the mask will remain complete for all angles in the FOV covered by the mask, albeit with less photons hitting the detector elements as the inclination increases. If we make the collimator blades high enough, there will be no partially-coded FOV, but the sensitivity of the instrument will decrease as the source incidence angles tilt away from the instrument axis. To ensure that no incident photons hit the detectors without passing through the mask (i.e.\ the PCFOV is really zero), it is easy to show (see Figures \ref{fig:fov} and \ref{fig:imp}) that the height of the collimator blades $h$, in an instrument with a simple geometry (e.g.\ square or circular), must be
\begin{equation}
h = \frac{2 d l}{M - D + 2 d}
\label{eq:coll}
\end{equation}
where $d$ is the distance between the collimator blades, $l$ is the PSDP-mask separation, $M$ is the mask linear size (the side or the diameter of the mask), and $D$ the PSDP linear size (side or diameter). The FCFOV in this case will have a HWZI (``half width at zero intensity'') of $\tan(d/h)$. In the extreme case in which the mask and the PSDP have the same size ($M = D$), equation \ref{eq:coll} gives $h = l$ and the HWZI  of the FCFOV is $ \tan(d/l)$. (Incidentally, it would make no sense to build such an instrument since it would produce no images and would be only a set of open and closed collimated detector cells.)

It is interesting to estimate the amount of shielding material that will have to be used in the two approaches. Since this amount is proportional to the area of the shielding walls for fixed thickness, in the first approach the amount will be approximately $\propto 4\times (D+M)l/2 = 2 l (D+M)$ for the reasonable assumption that $l \gg (M-D)/2$. In the second approach, we first note that the number of collimator blades in one dimension will clearly be $N_b = D/d +1$, including the external ones. The amount of shielding will then be $\propto 2 N_b h D = 4(D+d)lD/(M-D+2d) $. When the position resolution of the PSDP is reasonably good ($\Delta s \approx d \ll D$), this will be $\approx 4 l D^2 /(M-D)$, which means that the amount of shielding material for a mask with $M \approx 2D$, which is a fairly conventional configuration for a CAI instrument, will be about 33\% less in the collimator approach.

\section{The mask pattern}
\label{sec:pattern}

A major question in the design of CAI instruments is the choice of the pattern of cells in the mask that best suits the science requirements of the experiment and the practical implementation constraints. 

Since the mask in general has a larger area than the PSDP, only a fraction of its pattern will be cast over the PSDP by distant point sources, with in principle different shadowgrams for each one. If the mask is a cyclic repetition of a basic pattern wich has an area that fits inside the PSDP area, each source in the FCFOV will cast a shadowgram that will contain a {\it permutation\/} of the basic pattern, meaning that the source direction will be given by the $x$ and $y$ shifts of the basic pattern cast over the PSDP with respect to a source direction parallel to the instrument axis. If the mask is large in the sense that it contains several repetitions of  the basic pattern, there will be ambiguities in the source directions since the shadowgrams will repeat themselves over whole cycle permutations of the mask basic pattern. The best way to avoid that in the usual rectangular mask instruments is to use a basic pattern area that matches the detector surface and repeat it cyclically in the mask in a 2$\times2$ configuration, removing one row and one column at the edges of the mask. In this way, the mask shadow for any direction in the FCFOV will never repeat itself in the surface of the PSDP and it will always be a permutation of the basic pattern. Clearly, this scheme only works properly if the permutations themselves are always different than the original pattern, so any regular pattern (e.g.\ a chess board) will be a bad choice. It is noteworthy, however, that in some wide-field applications the mask is so large that this scheme is not possible, and the ambiguity problem has to be dealt with in conjunction with other instrumental constraints. One possible way to avoid the ambiguity is to use a rotating mask \citep{1984ITNS...31..771C, 2002RScI...73.3619B}. In this case, the center of rotation of the shadowgram in the PSDP will depend upon the source incoming direction and the ambiguity can be avoided by determining this center in the timing analysis of the recorded data. It is noteworthy that rotating masks can also be a good strategy to alternate exposures with mask and antimask if the mask pattern is chosen appropriately (see section \ref{sec:antimask}). 

An important point is that the geometry of the PSDP does not necessarily has to match the geometry of the mask pattern. For example, hexagonal mask patterns, such as the one used in Caltech's GRIP balloon experiment \citep{1984ITNS...31..771C}, were used so as to approximately match the circular surface of widely-used NaI(Tl) Anger cameras \citep{1985ITNS...32..129C} or other cylindrically-shaped position-sensitive detectors. However, \citet{1990TeseBRAGA} has shown that there is no loss neither in sensitivity nor in imaging properties if one uses a retangular mask pattern, provided any permutation of a full basic pattern of the mask is always projected onto the PSDP. This is due to the fact that the counts of basic-pattern spatially-equivalent pixels on the PSDP, that record statistically equivalent number of counts, can be added up and the results are then divided by the number of times these pixels are repeated on the PSDP surface. This clearly preserves the imaging properties of the pattern whereas the statistical fluctuations on the detector counts will correspond to the total counts over the whole detector, not only the counts in a basic-pattern equivalent area. As a consequence, the signal-to-noise ratio of the images will be determined by the total source and background counts, exactly as in experiments in which the PSDP area matches exactly the mask basic-pattern used.

The question of suitable mask patterns for CAI instruments has been studied in detail by many authors (see, for example, \citet{1995ExA.....6....1S} for a review). In general, the {\it open fraction\/} $f$ (the number of open cells over the total number of cells, or the open area over the total area of the basic pattern) has to be large enough ($f \gtrsim 0.3$) to provide high throughput (this is the essence of CAI). However, it cannot be too large, since in this case the number of ``dark'' regions on the PSDP for a given source will be too small and the measurement of the background against which the source signal will be evaluated will be statistically poor, degrading the signal-to-noise ratio of the reconstructed image. Therefore, an open fraction close to $1/2$ is in most applications a desirable property of the basic pattern. There have been important studies that address the maximization of the mask open fraction for specific applications (e.g. \citet{1994A&A...288..665I}), especially taking into account the difference in intensity between the intrinsic instrumental background and the diffuse radiation that comes through the aperture. 

\begin{figure}[!ht]
\centering
\includegraphics[width=0.45\hsize,angle=0]{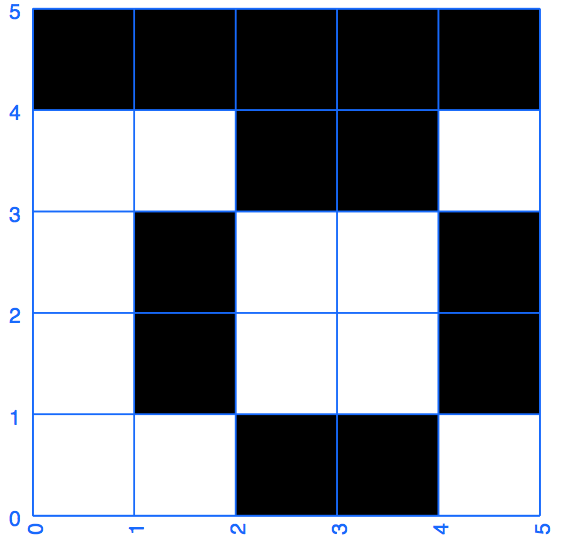}
\includegraphics[width=0.41\hsize,angle=0]{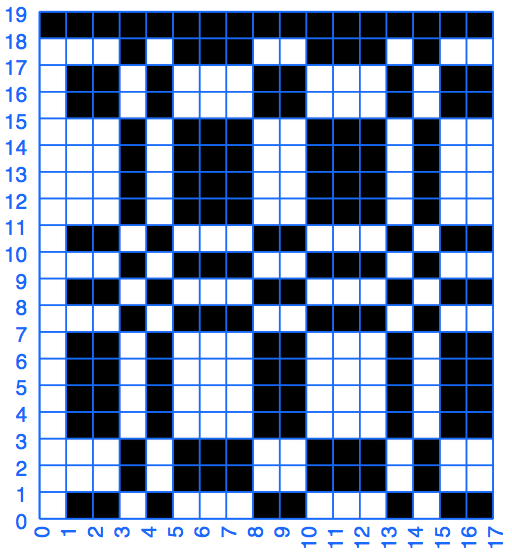}
\caption{{\it Left\/}: A 5x5 MURA pattern. {\it Right\/}: A 19x17 URA pattern.}
\label{fig:ura_mura}
\end{figure}

A class of widely used patterns comprises the so-called Uniformly Redundant Arrays (URAs) proposed in a seminal paper by \citet{1978ApOpt..17..337F}. They are ``uniformly redundant'' in the sense that the number of pairs of holes with a given separation in the pattern is the same for all possible separations, up to a maximum separation. The URAs have an open fraction of about $1/2$ and possess the desirable property of never repeating themselves in any permutation. In addition, every permutation of an URA is a very poor reproduction of the original pattern and all permutations are {\it equally poor} in the sense that they produce the same coefficient in the autocorrelation function, meaning that URAs enable images with no artifacts or ``intrinsic noise''. A two-dimensional URA is a rectangular pattern built with so-called twin-prime numbers (primes separated by 2). An interesting modification (the Modified URAs or MURAs) was introduced by \citet{1989ApOpt..28.4344G}; they noticed that a slight change in the decoding procedure (the inversion of one element in the decoding function that normally mimics the mask pattern) allows the use of square patterns with any prime number. The MURAs have exactly the same imaging properties as the URAs and provide a much wider choice of patterns to instrument designs. Later, \citet{2002RScI...73.3619B} noticed that a sub-class of MURAs have 90-degree symmetries, meaning that a 90-degree rotation of the mask produces an ``antimask'' with the exception of one single cell that remains closed. Imaging alternatively with mask and antimask can be an efficient way to subtract out systematic spatial variations in the background level across the detector plane (see section \ref{sec:antimask} and \citet{1991ExA.....2..101B}). Figure \ref{fig:ura_mura} shows a 5x5 MURA and a 19x17 URA patterns.

\begin{figure}[!ht]
\centering
\includegraphics[width=0.41\hsize,angle=0]{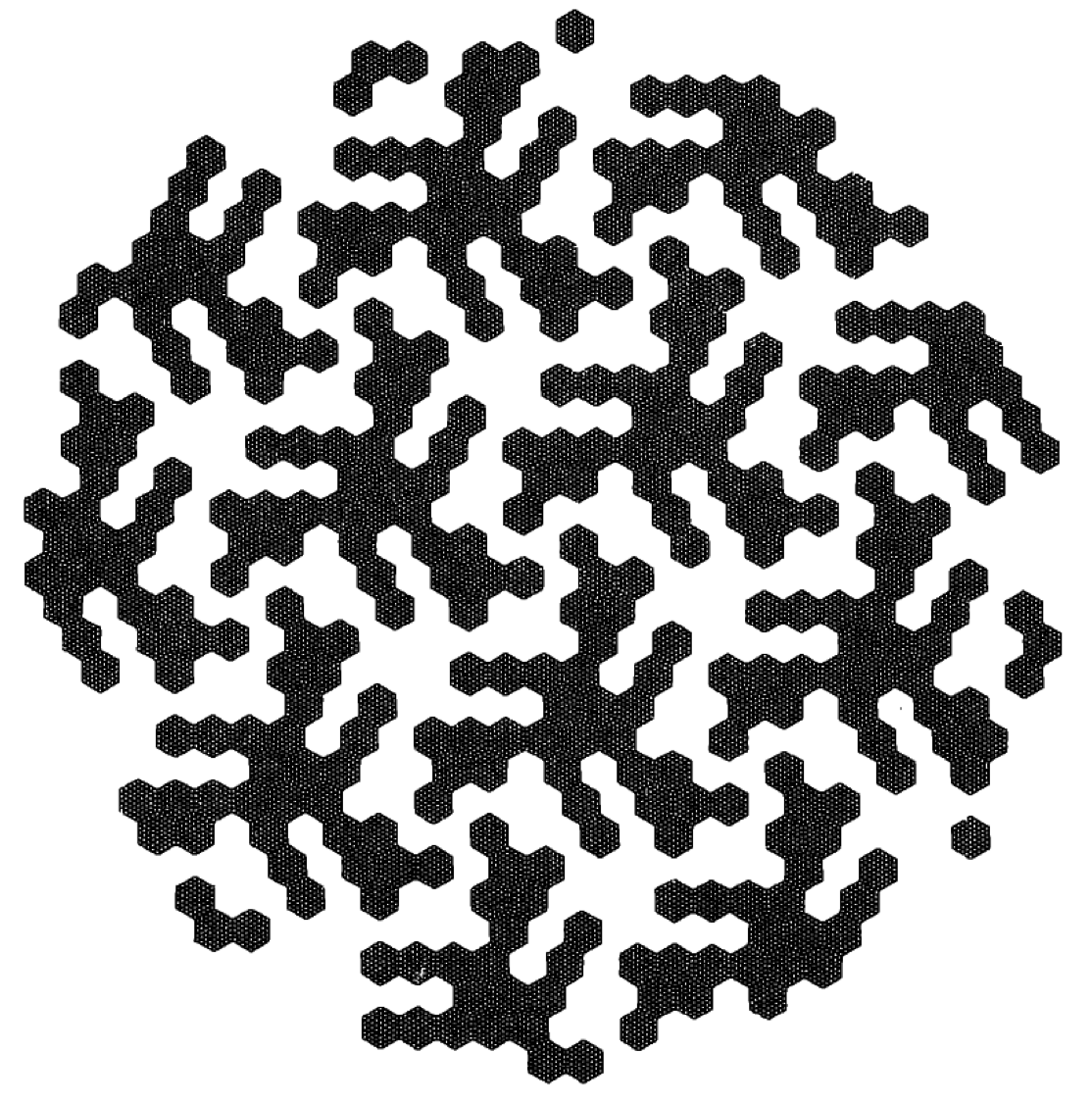}
\includegraphics[width=0.51\hsize,angle=0]{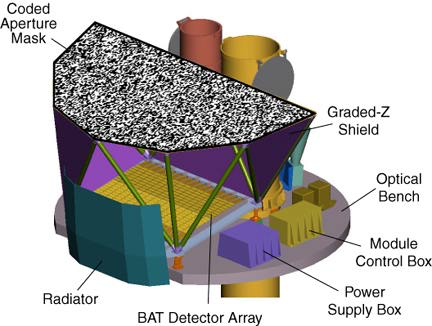}
\caption{{\it Left:} an hexagonal URA - HURA - order 67. These patterns belong to a special class URAS, the {\it skew-Hadamard\/} URAs, constructed on a hexagonal lattice, which have the interesting property of being antisymmetric upon rotations by 60$^\circ$ \citep{1985ICRC....3..295F}; see section \ref{sec:antimask}. {\it Right\/:} the random mask of 54k lead cells of the BAT/{\it Swift\/} telescope (swift.gsfc.nasa.gov).}
\label{fig:patterns}
\end{figure}

When the statistical fluctuations in the reconstructed image are large compared to the intrinsic noise of the pattern, the no-artifact property becomes less relevant. Random patterns with $f \approx 1/2$ have been used with good performance in several instruments (see Table \ref{tab:CAI}), especially when the number of elements in the pattern is large ($\gg 1$) and imaging in the PCFOV is important. In applications in which the number of mask elements needs to be small (usually in $\gamma$ rays), it is important that the autocorrelation function of the pattern  approaches a delta function so that the reconstructed image has no artifacts. Figure \ref{fig:patterns} show other examples of coded-mask patterns.

\section{Image formation}
\label{sec:formation}

It is instrumental to divide the FCFOV in {\it skybins}. The central skybin, associated to the axial direction, can be defined by the solid angle defined by a mask cell as viewed from the detector plane directly below it. For a PSDP with a spatial resolution that approximately matches the mask cell size ($\Delta s \approx m$), we can define statistically independent pixels with an approximately size of $m\times m$. Let us define these pixels exactly under the mask cells in the direction of the instrument's axis. The intensity map in the PSDP can be defined by an array $D$  whose elements are the intensities (essentially counts) in each one of the pixels defined above.

In this approach, the central angle of the next skybin in a certain direction following a given row of cells in the mask will be given by $\theta =\tan^{-1} m/l$, since the distance to the next cell center is $m$. In this skybin, the shadows of the mask elements will clearly be cast over the neighboring pixels on the PSDP in the opposite direction. The shadowgram for this skybin, neglecting the effects of the thickness of the mask cells, will have the same size as the one cast by the central skybin. However, the number of counts in each pixel will be diminished by $\cos\theta$ (see Figure \ref{fig:fov}) due to the projected area.

Proceeding in this way we can divide the entire FOV of CAI instrument in a grid of skybins whose central angles are given by
\begin{equation}
\theta =\tan^{-1} \xi/l
\end{equation}
where $\xi$ is the distance from the specific cell to point O shown in Figure \ref{fig:fov}. If $\xi < \chi$, where $\chi = (M - D)/2$, then $\theta$ is inside the FCFOV, whereas if $\chi < \xi < \chi + D$, then $\theta$ is in the PCFOV. For $ \xi > \chi + D$, which would indicate a point beyond the edge of the mask, the flux is completely uncoded by the mask (in general, the shielding system absorbs incoming radiation from these directions).

\begin{figure}[!ht]
\centering
\includegraphics[width=0.35\hsize,angle=0]{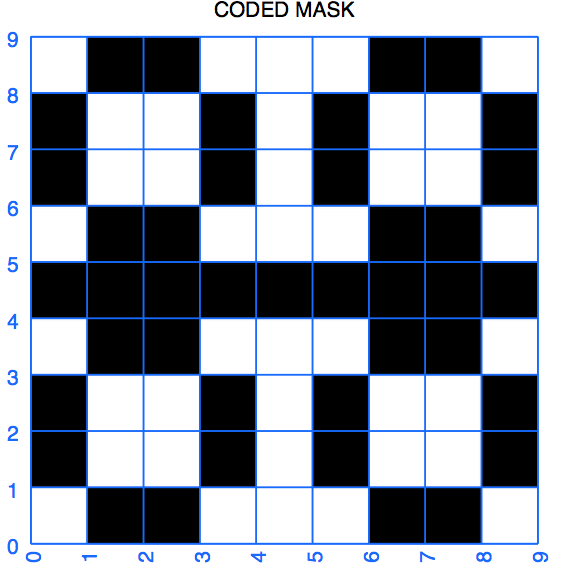}
\includegraphics[width=0.55\hsize,angle=0]{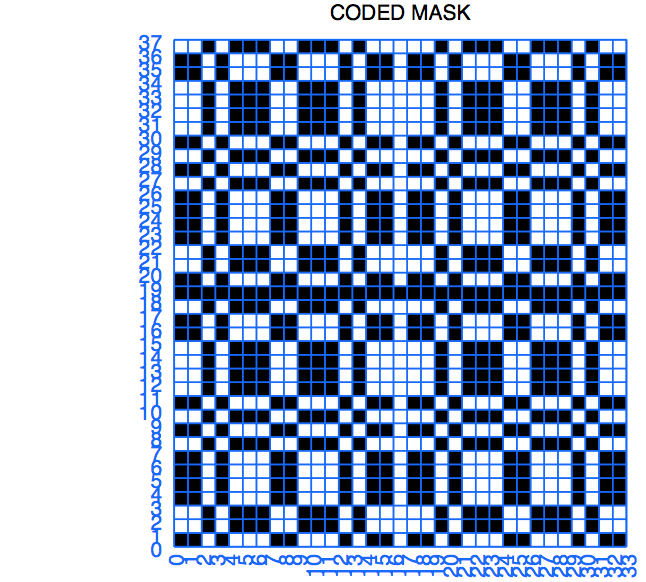}
\caption{{\it Left\/}: A 5x5 MURA-based mask. {\it Right\/}: A 19x17 URA-based mask. Both masks are 2x2 cyclic repetitions of the basic pattern, minus one row and one column.}
\label{fig:mask_ura_mura}
\end{figure}

If the mask is a $2\times 2$ cyclic repetition (minus 1 row and 1 column) of a basic pattern of $p\times q$ elements that matches the detector area, each skybin of the FCFOV will cast a shadowgram that is a different permutation of the basic pattern over the PSDP, and all permutations of it will be present. For example, convenient masks for the two patterns shown in Figure \ref{fig:ura_mura} are shown in Figure \ref{fig:mask_ura_mura}.

The intensity of each skybin can be recovered by a matching process. If we represent the mask basic pattern by an array $M_{ij}$ whose elements are 1 for the open cells and 0 for the closed ones, what gets imprinted in the PSDP can be described mathematically as a {\it correlation\/}:
\begin{equation}
D = S\otimes M + B,
\end{equation}
where $\otimes$ is the correlation operator. In a digital implementation, this can be written as
\begin{equation}
D_{ij} = \sum_{k=0}^{p-1} \sum_{l=0}^{q-1} S(k,l) M((k+i)\, {\rm mod}\, p, (l+j)\, {\rm mod}\, q) + B_{ij}\;\;\;\; ; \;\;i=0\ldots p-1, j=0\ldots q-1
\label{eq:corr}
\end{equation}
where the elements of $S$ are the intensities for sources in each skybin (counts per cell area) multiplied by the $\cos\theta$ factor, and $B$ is an aperture-independent (instrumental) background (also in counts/cell). An approximately angularly-constant diffuse sky background that enters through the aperture will also be independent of the mask pattern, since all sky-bins will have the same intensity. This will add a DC-level to the count map, so its effect can be added to $B$. However, it is important to make a distinction between the instrumental and sky backgrounds, since the latter scales with the open fraction of the mask and the former does not. The relative intensities of these two types of background are relevant factors to be taken into account when designing a CAI instrument and its mask pattern (see e.g.  \citet{1994A&A...288..665I} for a discussion about this point.)

If the total FOV is completely coded, the elements of $D$ will have the combined contributions of one permuted basic-pattern shadowgram per skybin. If, however, there is a partially-coded FOV, each element in $D$ will have contributions from all ``equivalent" shadowgrams of those partially-coded skybins that cast the same permuted shadowgram. These incomplete shadowgrams will produce artifacts that could appear as fake or ``ghost'' peaks in the final image. This is especially undesirable if there are strong sources in the PCFOV because the ghost peaks could be confused with real sources.

In order to obtain S from D, several approaches and algorithms have been discussed in the literature (see \citet{2003astro.ph..2354S} for a review) and used in specific instruments with specific detector systems and mask patterns \citep{1992NIMPA.311..585H, 1987Ap&SS.136..337S, 2010A&A...519A.107K, 2005ApJ...618..856C, 2002ApOpt..41..332H, 1996A&AS..120..579R}. A widely used reconstruction method, described in this paper, is the correlation of the recorded data with a suitable decoding function that in general mimics the mask pattern. In these cases Fast Fourier Transforms (FFT) have been extensively used to speed-up the reconstruction process, since the correlation operation in Fourier space becomes a multiplication. In what follows, the focus is on the properties of the reconstructed images rather than on the particular procedure to acquire it. Since $D = S\otimes M + B$, 
\begin{equation}
D\otimes G = (S\otimes M)\otimes G + B\otimes G = S\otimes (M\otimes G) + B\otimes G
\end{equation}
where $G$ is the so-called {\it decoding function}. If $M\otimes G$ is a two-dimensional discrete $\delta$-function, the reconstructed image $\hat{S}$ is given by 
\begin{equation}
\hat{S} = D\otimes G - B\otimes G,
\label{eq:rec}
\end{equation}
Therefore, the intensities of each skybin (the image) are obtained by a cross-correlation of the count distribution in the PSDP with the decoding function and the image has no artifacts (sidelobes). If the background is uniform, the second term on the right side of equation \ref{eq:rec} just adds a DC-level to the reconstructed image. If the background is known by an independent measurement, $\hat{S} = (D - B)\otimes G$ (see section \ref{sec:antimask}). In both cases, the source map is reconstructed ``perfectly'' in the sense that it is free of intrinsic noise.

If $M$ is such that its autocorrelation function is a $\delta$-function, we can choose $G=M$ and the decoding function is simply given by the mask basic pattern. There are several classes of patterns that approach this property, the most famous of which being the URAs \citep{1978ApOpt..17..337F} already mentioned in section 5. The URAs are built by ``twin'' prime numbers $p$ and $q$, $p-q=2$. The autocorrelation function of an URA has a peak with value $(pq+1)/2$ (which is precisely the number of open cells in the pattern) and a flat ``DC-level'' with half that value. If we define a $G$ function simply replacing the ``0''s in $M$ by  ``-1''s, we get $M\otimes G$ with the same peak and no DC-level. Therefore, by using a unitary array $G$ whose elements mimic the mask basic pattern and are either 1 (for open cells) or -1 (for closed ones) we can produce an image of the FOV with no artifacts and zero base level (off-peak regions). For MURAs, the autocorrelation of the pattern is not a $\delta$-function, but if we invert one element of $G$ \citep{1989ApOpt..28.4344G, 2002RScI...73.3619B}, then $M\otimes G$ is a $\delta$-function and we get the same properties in the reconstructed images; the only difference, which is important only for masks with a small number of cells, is that the peak (and the number of open cells) is now  $(pq-1)/2$.

\section{Masks and antimasks}
\label{sec:antimask}

One relevant aspect of CAI instruments that has to be carefully taken into consideration is the uniformity of the position-sensitive detector plane. This depends on the nature of the detectors and on the geometry of shielding materials and structural parts surrounding the PSDP. 

Sensitivity differences among PSDP pixels can be corrected for by standard flat-fielding procedures, such as the ones used in CCD optical sensors, in which multiplicative factors are applied to each energy-calibrated pixel in order to correct the count rates with respect to a mean value, for each energy range of interest. The factors are determined by carrying out an equal illumination of all PSDP pixels by a bright source, prior to the actual observation of a source field by the instrument.

On the other hand, spatial variations of the background level across the PSDP have to be dealt with differently, since in an observation of a source field one cannot in principle separate the additive contribution of source counts in a pixel from an intrinsic spatial non-uniformity of the background rate. In a CAI instrument operation, source and background are observed simultaneously, so normally there is no independent determination of the background in each pixel to be subtracted from the source. This is actually a major advantage of CAI instruments over non-imaging collimated instruments, since the latter have to spend precious time observing adjacent fields to measure the background that could change both spatially (different sky region and/or local observing direction) and in time. Moreover, the background of space instruments are of complex origin and can vary in short time scales \citep{1985NIMPA.239..324G}. One efficient way to eliminate this problem is to break down the integration time of an observation of a field in two halves, one with the normal mask and another one with an ``antimask", an inverted pattern of opaque and transparent cells (see \citet{1991ExA.....2..101B}). From equation \ref{eq:rec}, the two images $\hat{S}_m$ and $\hat{S}_a$ will be given by 
\begin{equation}
 \hat{S}_m = (D_m - B)\otimes G\;\; ; \;\; \hat{S}_a = (D_a - B)\otimes \bar{G},
\end{equation}
where $D_m$ is the count distribution obtained with mask and $D_a$ the one with the antimask. $\bar{G}$ is the decoding function corresponding to the antimask, thus $\bar{G} = -G$. Therefore 
\begin{equation}
\hat{S}_m + \hat{S}_a = (D_m - B)\otimes G + (D_a - B)\otimes\bar{G} =  (D_m - D_a)\otimes G
\end{equation}
which completely eliminates the background. Now, $D_m$ and $D_a$ represent the same distribution of events (within statistics), but in alternate pixels, so $D_m - D_a$ will have positive and negative values. The summed image has the same statistical properties as an image with a mask for the whole integration time. 

The main problem of using antimasks, besides the addition of possibly significant mass to the instrument, is the mechanical complexity to alternate mask and antimask mountings in the instrument FOV. However, as mentioned in section \ref{sec:pattern}, there are mask patterns that produce antimasks of themselves (with the exception of a singe cell) by a simple rotation, which mitigates the problem significantly. Examples include a class of MURAs \citep{1991ExA.....2..101B} and hexagonal URAs \citep{1984ITNS...31..771C}.

\begin{figure}[!ht]
\centering
\includegraphics[width=0.44\hsize]{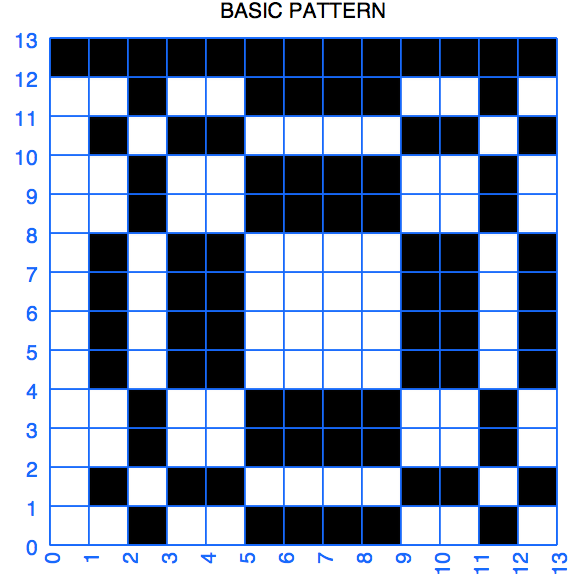}
\includegraphics[width=0.515\hsize]{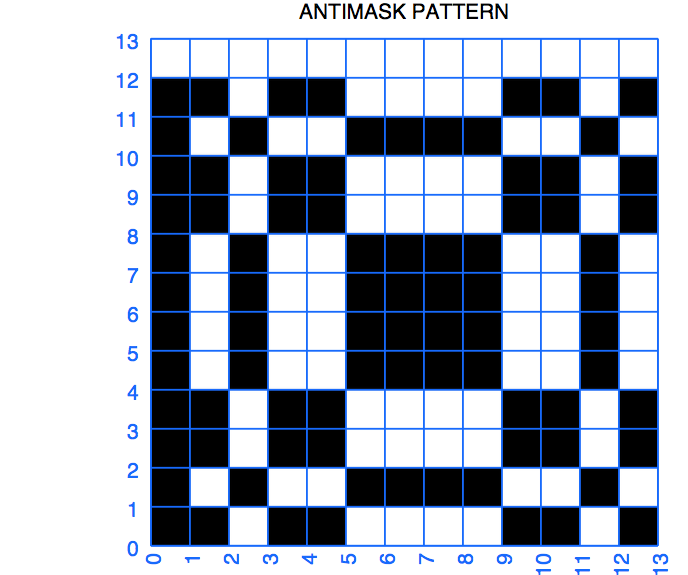}
\caption{The MURA 13x13 mask pattern and its ``antimask'' pattern.}
\label{fig:antimask}
\end{figure}

In order to demonstrate the method, I have run Monte Carlo simulations to show the difference in signal-to-noise ratios of point source images when a systematic background variation occurs. In the following example, I used a 13x13 MURA mask and its antimask, shown in Figure \ref{fig:antimask}.

\begin{figure}[!ht]
\centering
\includegraphics[width=0.45\hsize]{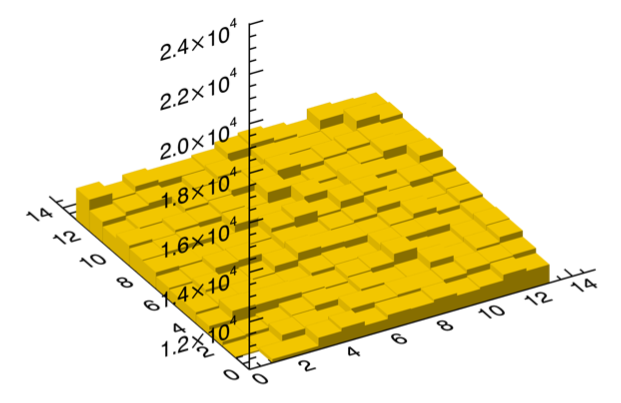}
\caption{A simulated background distribution on a 13x13-pixel PSDP with a 20\% Poissonian ramp variation above a 10k mean count value at each pixel. Only the top of the distribution is shown.}
\label{fig:bkg}
\end{figure}

\begin{figure}[!ht]
\centering
\includegraphics[width=0.4\hsize]{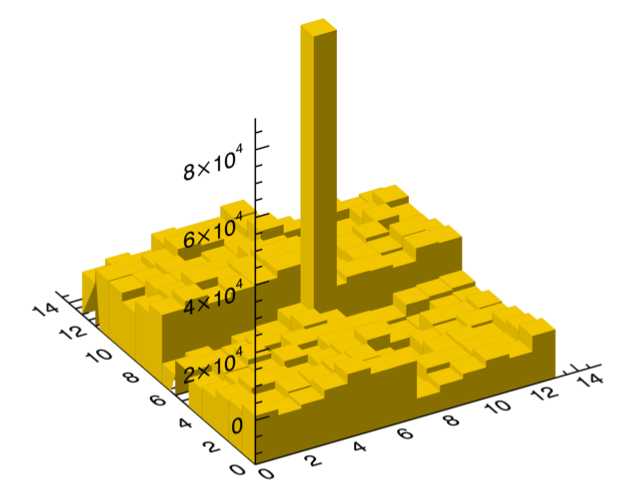}
\includegraphics[width=0.4\hsize]{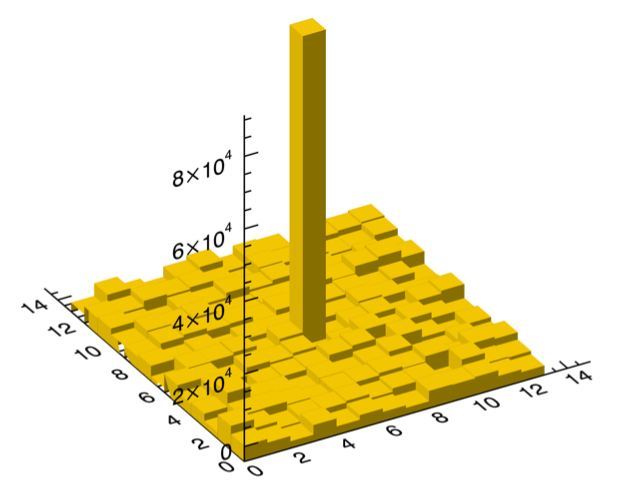}
\caption{simulated reconstructed images with a 13x13 MURA mask, 10k background counts per pixel and 1k source counts per illuminated pixel. {\it Left\/}: The image produced using only the basic pattern shows artifacts due to background systematic spatial non-uniformity across the PSDP. {\it Right\/}: The image produced with the mask/antimask technique completely eliminates background systematics and the peak has the correct statistical significance.}
\label{fig:sim_images}
\end{figure}

I ran a simulation with a mean of 10k background counts in each pixel as to have 1\% statistical fluctuations (Poisson statistics). Then I introduce a systematic count ramp with 20\% maximum variation toward one of the corners of the PSDP (Figure \ref{fig:bkg}). The simulated source, placed at the centre of the FOV, has 1k counts per illuminated pixel, so the background is ten times stronger than the source in each pixel. The reconstructed images with the mask show strong artifacts due to the nonuniform background, whereas the images produced with the mask/antimask technique completely eliminate the artifacts and have signal-to-noise ratios (SNRs) that are statistically equivalent to the theoretical one given by Poisson statistics, which in this particular case is $\sim 62$ (see section \ref{sec:snr}). The mask-only images have an average SNR of $\sim 16$. Figure \ref{fig:sim_images} shows the two images.

\section{Signal-to-noise ratio and sensitivity}
\label{sec:snr}

With the knowledge of the expected background levels of a coded-mask imager as a function of energy, we can calculate the sensitivity of a URA-like coded mask instrument in terms of the minimum detectable flux at a particular statistical significance. From equations \ref{eq:corr} and \ref{eq:rec}, using the unitarity of the $G$ function, the variance of each pixel of the reconstructed image can be calculated by error propagation, using Poisson statistics \citep{1989ApOpt..28.4344G}:
\begin{equation}
\sigma^2 = N_S + N_T 
\end{equation}
where $N_S$ is the net source counts for a particular skybin and $N_T$ is the total number of counts in the PSDP, all of them for the same integration time in a particular energy range.  If there are multiple sources in the field, each skybin that contains a source ($S_n$ counts per cell for the $n$-th source) will produce a peak in the reconstructed image that has a value of $\hat{S}_n = N_H S_n$, where $N_H$ is the number of holes in the pattern; this skybin will not contribute to the other pixels of $\hat{S}$. 
Therefore, the signal-to-noise-ratio (SNR) of each peak of the reconstructed image  is given by
\begin{equation}
S\!N\!R_n = \frac{N_H S_n}{\sqrt {N_S + N_T}}\;\;\; ; n=1, 2, \ldots 
\label{eq:snr}
\end{equation}
Now, we have that
\begin{equation}
N_T = N_H \sum_n S_n +  N_C B, 
\end{equation}
where the summation encompasses all sources and $N_C$ is the total number of cells in the basic pattern. Typically (e.g.\ URAs), $N_C \approx 2 N_H$ and we can write
\begin{equation}
S\!N\!R_n = \frac{N_H S_n}{\sqrt{N_H S_n + N_H \sum S_n + 2 N_H B}} = \frac{\sqrt{N_H} S_n}{\sqrt{S_n + \sum S_n + 2 B}}
\label{eq:snr2}
\end{equation}

A crucial parameter in the design of a CAI instrument is its sensitivity for one point source in the FCFOV. Equation \ref{eq:snr2} for one single source becomes
\begin{equation}
S\!N\!R = \frac{N_H S}{\sqrt{ N_H S + 2 N_H B}}.
\end{equation}
If we are interested in a {\it minimum detectable flux\/} for a point source, it is reasonable to assume that $S \ll 2B$. Now, if $F$ is the source flux in, for example, photons cm$^{-2}$s$^{-1}$keV$^{-1}$, then 
\begin{equation}
F = \frac{N_H S}{A_{\rm eff} \, T\, \Delta E},
\end{equation}
where  $A_{\rm eff}$ is the effective area in cm$^2$ (which takes into account the open fraction of the mask and other parameters like detector efficiency), $T$ is the integration time in seconds and $\Delta E$ is the energy range under consideration (in keV). Similarly, the background flux $F_B$ in counts cm$^{-2}$s$^{-1}$keV$^{-1}$ is given by
\begin{equation}
F_B =  \frac{2 N_H B}{A_{\rm geo} \, T\, \Delta E}.
\end{equation}
where $A_{\rm geo}$ is the geometrical area (in cm$^2$) of the whole detector plane. 

The minimum source flux that will be detectable at a level of $N_{\sigma} \equiv S\!N\!R$ will then be
\begin{equation}
F_{\rm min} = \frac{N_{\sigma}}{A_{\rm eff}}\; \sqrt{\frac{B \, A_{\rm geo}}{\Delta E\; T}} \;\; {\rm photons\; cm^2 s^{-1} keV^{-1}}.
\label{eq:sens}
\end{equation}
We note again that this equation is only valid for URA-like patterns for which $N_C \approx 2 N_H$, meaning that the mask open fraction is $\sim 1/2$. See \citet{2012NIMPA.669...22S} for specific cases of CAI signal-to-noise ratios, including different patterns and open fractions.

It is interesting to compare the sensitivitiy of a coded-aperture camera with a similar instrument which has an open aperture (a similar camera without the mask). There is a factor of 2 advantage in $A_{\rm eff}$ for the open camera due to the absence of a mask, but the instrument would have to observe only background for approximately the same amount of time as the on-source time in order to have the same statistics. According to equation \ref{eq:sens}, the coded-mask camera would be less sensitive by a factor $\sqrt{2}$ for an observation of only one point source. However, the coded-mask instrument can take fluxes and spectra of several objects at the same time, albeit with relatively less SNR due to the effect of the other sources in the noise term (see equation \ref{eq:snr2}). In addition, the fact that the coded-aperture camera measures source and background simultaneously and in the same sky field can be very important if there are significant time and spatial variations on the background level. 
 
 \section{Use of perpendicular one-dimensional coded aperture cameras}
 \label{sec:onedim}
 
Over the last two decades, with the advent of new detector technologies, imaging instruments have been developed with the combination of two perpendicular one-dimensional coded-mask cameras. The advantage of this scheme comes from the fact that different kinds of detectors with one-dimensional spatial determination capability can achieve high sensitivities with very fine spatial resolution, and the combination of two 1D position distributions can produce high-resolution 2D images. 

For example, the Wide-field X-ray Monitor (WXM) \citep{1995Ap&SS.231..463Y}  onboard the {\it HETE-2\/} satellite \citep{2003AIPC..662....3R} used two 1D position-sensitive proportional counters (PSPC) that achieved 1D position resolutions of $\sim 0.7$\ts mm at 8\ts keV. The use of the two cameras allowed GRB positions to be obtained with 10 arcmin accuracy.

Another interesting instrument that applies this approach is the SuperAGILE hard X-ray imager \citep{2007NIMPA.581..728F} on the Italian {\it AGILE} mission \citep{2008NIMPA.588...52T, 2009A&A...502..995T}, launched in 2007. The instrument uses silicon microstrip detectors in the 15-45\ts keV range, with the combination of the two crossed one-dimensional cameras providing images with a on-axis angular resolution of 6 arcmin over a FOV $> 1$\ts sr. 

\begin{figure}[!ht]
\centering
\includegraphics[width=0.4\hsize]{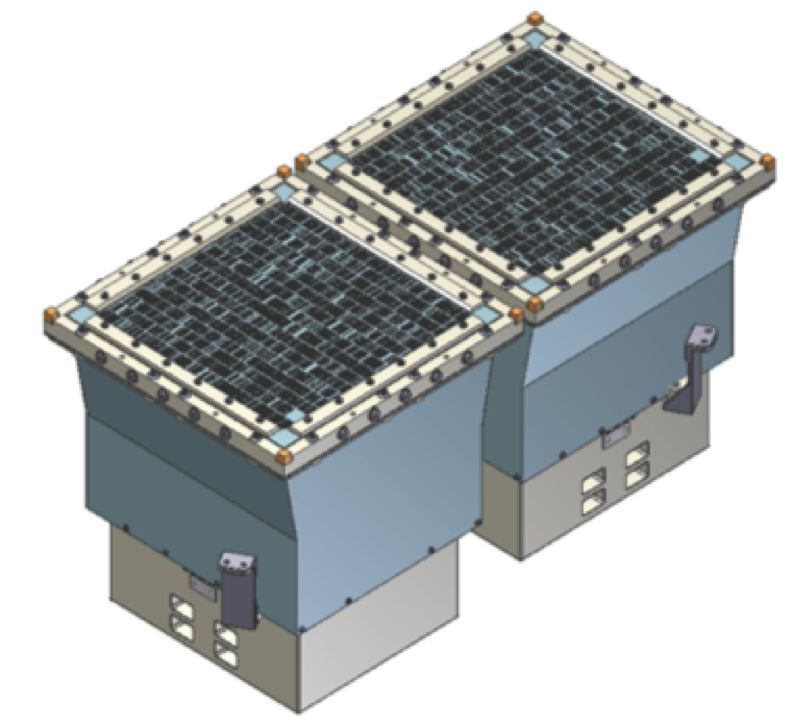}
\caption{A pair of Wide Field Monitors of {\it eXTP\/}, arranged orthogonally; from \citet{2018SPIE10699E..48H}.}
\label{fig:WFM_cameras}
\end{figure}

The same approach is planned to be used with large-area silicon drift detectors in the proposed missions {\it Strobe-X\/} \citep{2017ResPh...7.3704W} and {\it eXTP\/} \citep{2019SCPMA..6229502Z,2019SCPMA..6229506I}.  {\it Strobe-X\/} (Spectroscopic Time-Resolving Observatory for Broadband Energy X-rays) is a NASA Astrophysics Probes concept mission devoted to probe strong gravity from stellar mass to supermassive black holes and ultradense matter with unprecedented effective area and high time resolution. {\it eXTP\/} (enhanced X-ray Timing and Polarimetry) is an approved Chinese-European mission to be launched in the mid-2020s with 4 instrument packages (including X-ray polarization measurements). Its Wide Field Monitor (WFM) will include 3 pairs of coded-mask cameras covering 4 sr at a sensitivity of 4 mCrab for an exposure of 1 day in the 2-50 keV range. The angular resolution will be a few arcmin. In Figure \ref{fig:WFM_cameras} we show a pair of WFM cameras arranged orthogonaly \citep{2018SPIE10699E..48H}. 

\begin{figure}[!ht]
\centering
\includegraphics[width=0.33\hsize]{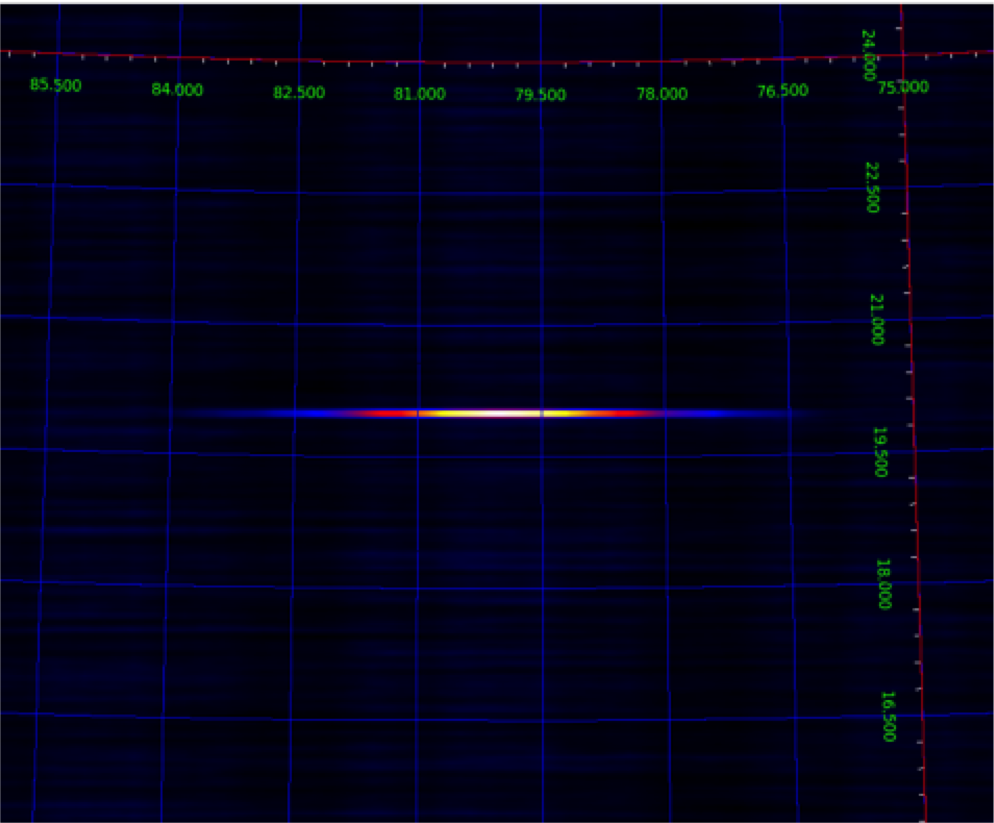}
\includegraphics[width=0.32\hsize]{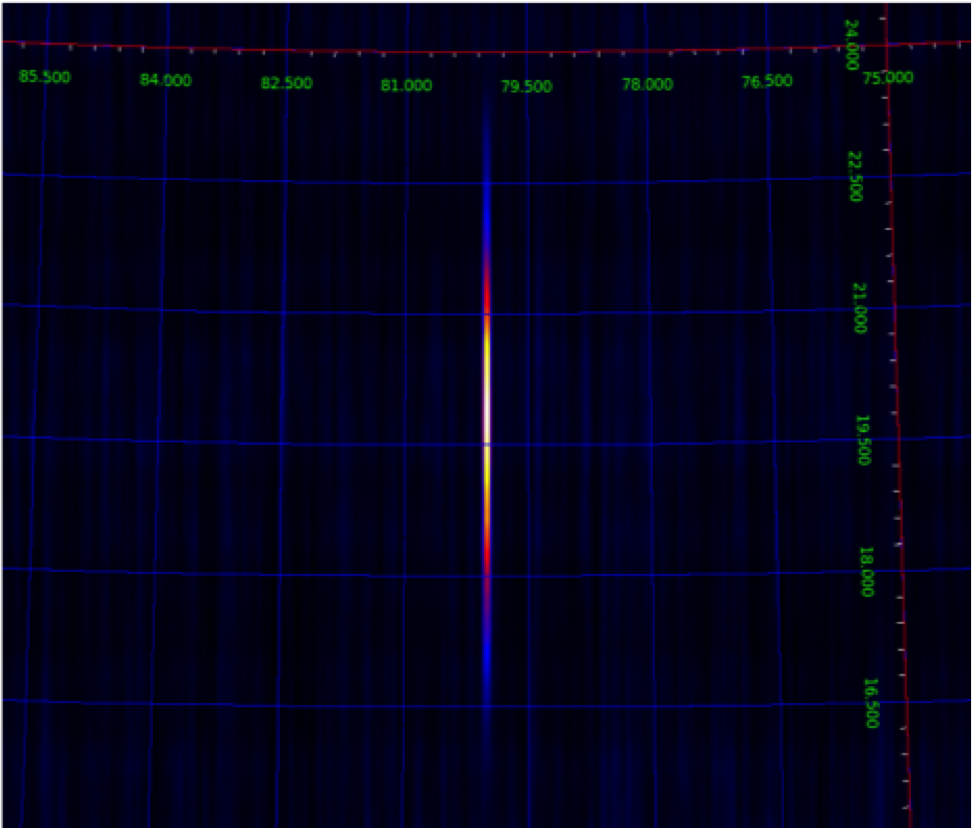}
\includegraphics[width=0.305\hsize]{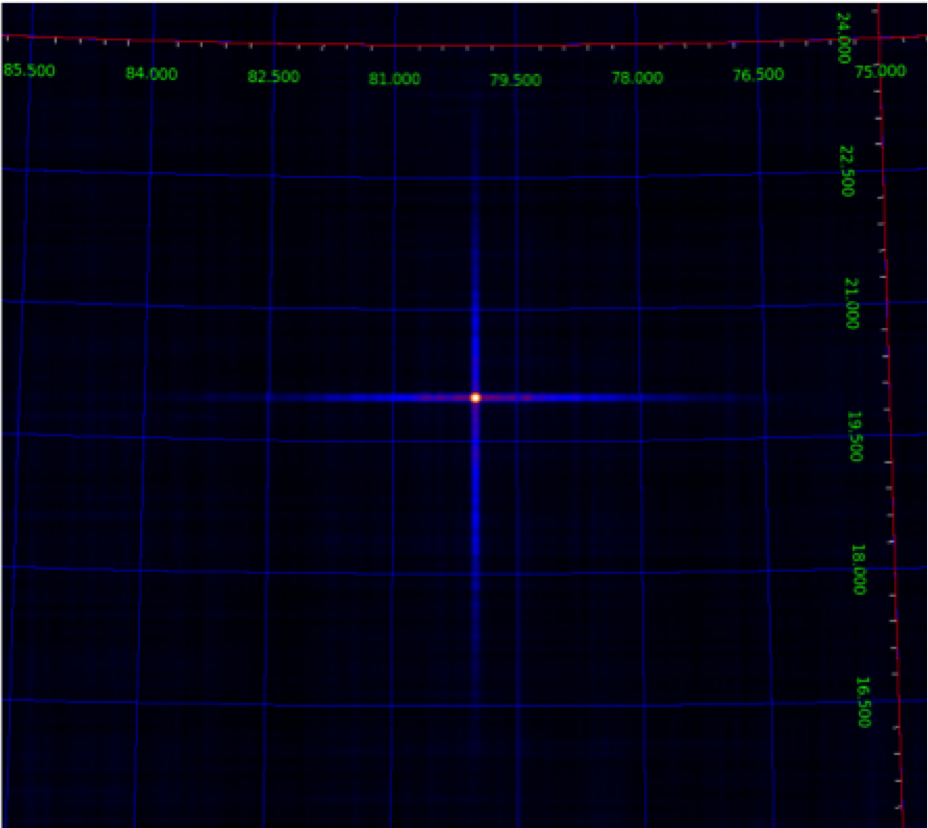}
\caption{WFM/{\it eXTP\/} simulated images of a central point source. The first two panels are ``1.5D'' images from two ortogonal coded-mask cameras. The third panel shows the combined 2D image of the field; from \citet{2018SPIE10699E..48H}.}
\label{fig:WFM_images}
\end{figure}

The position- sensitive Silicon Drift Detectors (SDD) \citep{1984NIMPR.225..608G} used in the WFMs provide accurate positions in one direction but only coarse positions in the perpendicular one. Using a coded mask with cell sizes that approximately match the position resolutions in both directions, one can produce ``1.5D'' positions of sky sources. Combining the two perpendicularly-oriented co-aligned cameras, a map with accurate 2D positions can be produced. Figure \ref{fig:WFM_images} shows a simulated image built by this procedure. 

\section{Contributions of coded-mask imagers to X-ray astronomy}
\label{sec:contrib}

Over the past decades, several coded-mask instruments have made significant discoveries and obtained important results in high-energy astrophysics. Table \ref{tab:CAI} shows the main parameters of some of the most important missions that used coded-aperture instruments onboard. In this section I comment on some of the most exciting results obtained by some of these missions.

\begin{table*}[!ht]
\small
\centering
\caption{Selected 2D coded-mask X-ray satellite experiments. PSPC: Position-Sensitive Proportional Counter; HPGe: Hyper-Pure Germanium; HPIMGC: High Pressure Imaging Microstrip Gas Chamber. The angular resolutions and FOVs are given in FWHM. Source: heasarc.nasa.gov}
\label{tab:CAI}
\begin{tabular}{|l|l|l|l|l|l|l|l|}
\hline
Launch & Payload & Satellite & Detector & Mask & Energy & Ang. & FOV \\
year & & & type & type & range (keV) & resol. & \\
\hline
1985 & XRT & Spacelab-2& PSPC & URA & 2.5-25\ & 3'-12'& $6^\circ$\\
1989 & SIGMA & GRANAT & Anger C. & URA & 30-1300\ & 15'  & $5^\circ$\\
1989 & ART-P & GRANAT &  PSPC & URA & 4-60\ & 5' &1.8$^\circ$ \\
1996 & WFC & Beppo-SAX & PSPC & pseudo-ran. & 2-30 & 5' & $20^\circ$ \\
2002 & SPI & INTEGRAL & HPGe & HURA & 18-8000 & $ 2.5^\circ$ & $ 14^\circ $ \\
2002 & IBIS & INTEGRAL & CdTe/CsI & MURA & 15-10000 & 12' & $ 8.3^\circ $ \\
2002 & JEM-X & INTEGRAL &  HPIMGC & HURA &  3-35\ & 3'  & $7.5^\circ$\\
2000 & WXM & HETE-2 & PSPC & random & 2-25 & 11' & $45^\circ$ \\
2004 & BAT & Swift &  CdZnTe & random  & 15-150 & 17' & 1.4 sr\\  
2015 & CZTI & ASTROSAT & CdZnTe & URA & 10-150 & 8' & 6$^\circ$ \\
\hline
\end{tabular}
\end{table*}

\begin{figure}[!ht]
\centering
\includegraphics[width=0.48\hsize]{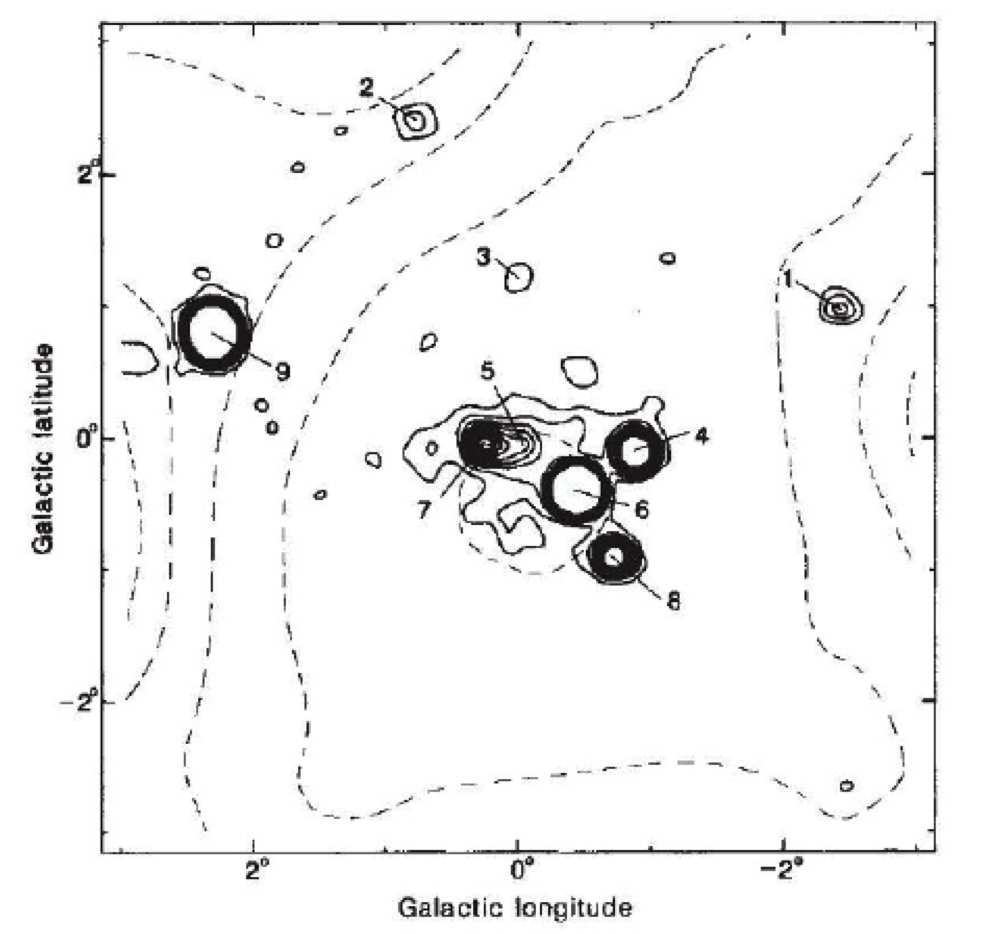}
\includegraphics[width=0.47\hsize]{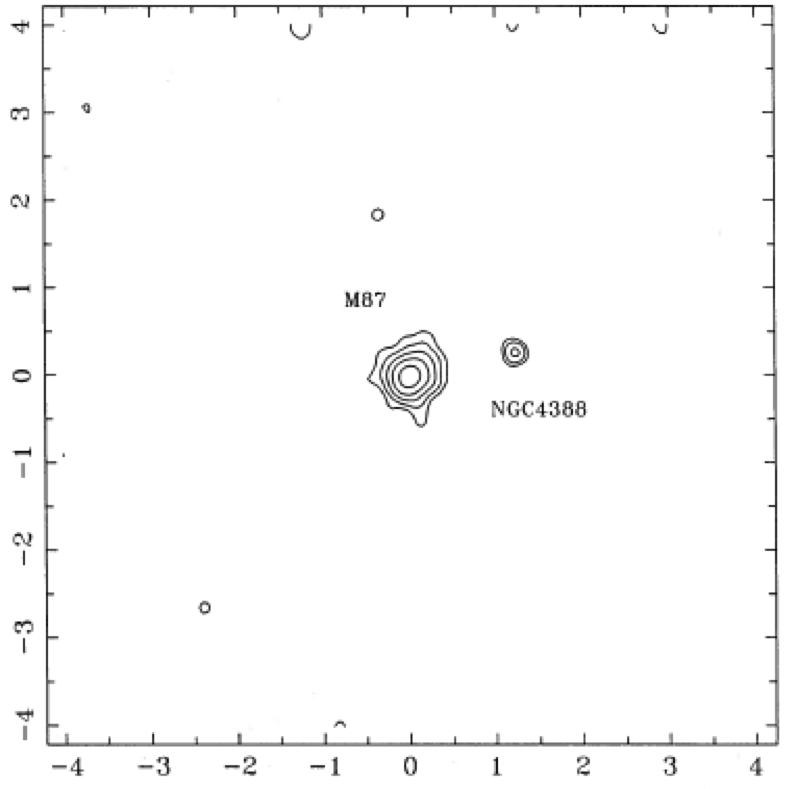}
\caption{Two astrophysical coded-mask images taken from the XRT telescope on {\it Spacelab-2\/}. {\it Left\/}: A GC image from 3 to 30\ts keV, showing weak emission from the center itself (Sgr\ts A$^\ast$ -- source 5) and strongest emission from GX\ts 3+1 (source 9); from \citet{1987Natur.330..544S}. {\it Right\/}: a $8^\circ\times 8^\circ$ image of the Virgo cluster from 2.0 to 9.6\ts keV, showing M87 and the Seyfert galaxy NGC\ts 4388; from \citet{1990MNRAS.242..262H}.}
\label{fig:xrt}
\end{figure}

The SL2-XRT coded-mask experiment on {\it Spacelab-2\/} took the first hard X-ray images of the Galactic Center (CG) region with a resolution of a few arcmin \citep{1987Natur.330..544S}. The images showed several new X-ray point sources and were able to isolate the emission from Sgr\ts A$^{\!\star}$. XRT also imaged the Virgo cluster \citep{1990MNRAS.242..262H}, showing that a significant fraction of the hard X-ray emission previously reported from the cluster comes from NGC\ts 4388, a Seyfert 1 galaxy. Figure \ref{fig:xrt} shows two XRT images. It is noteworthy that the XRT consisted of two units with different mask designs (yielding different angular resolutions) in otherwise identical instruments. The instrument with poorer resolution provided higher sensitivity to weak diffuse emission, which in this case was observed extending to about $1^\circ$ from the Galactic center (Figure \ref{fig:xrt}, left). 

\begin{figure}[!ht]
\centering
\includegraphics[width=0.65\hsize]{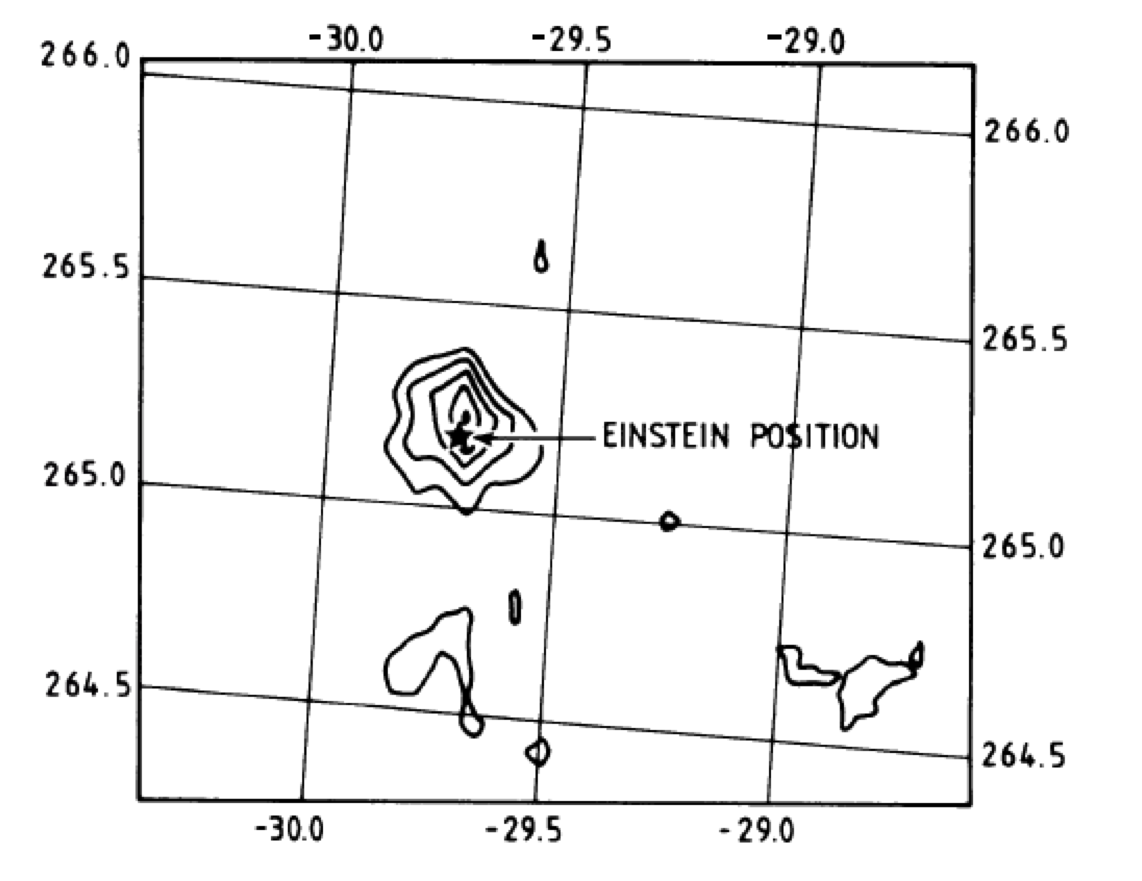}
\caption{A 330-570\ts keV Galactic Center image taken by the SIGMA coded-mask telescope onboard {\it GRANAT\/}. The coordinates are declination and right ascension. The image shows strong emission from 1E\ts 1740.7-2942 (Einstein source) probably associated with the 511\ts keV positron annihilation line; from \citet{1991ApJ...383L..45B}}
\label{fig:SIGMA}
\end{figure}

The French SIGMA coded-mask telescope onboard the Russian {\it GRANAT\/} satellite \citep{1991AdSpR..11..289P} obtained hard X-ray and gamma-ray images of a variety of source fields and yielded extremely important results. As an example, in Figure \ref{fig:SIGMA} we show an image of the GC region \citep{1991ApJ...383L..45B} in which the GC 511\ts keV $e^- \!\!-\! e^+$  annihilation feature was localized and shown to come from the microquasar 1E\ts 1740.7-2942.

\begin{figure}[!ht]
\centering
\includegraphics[width=0.75\hsize]{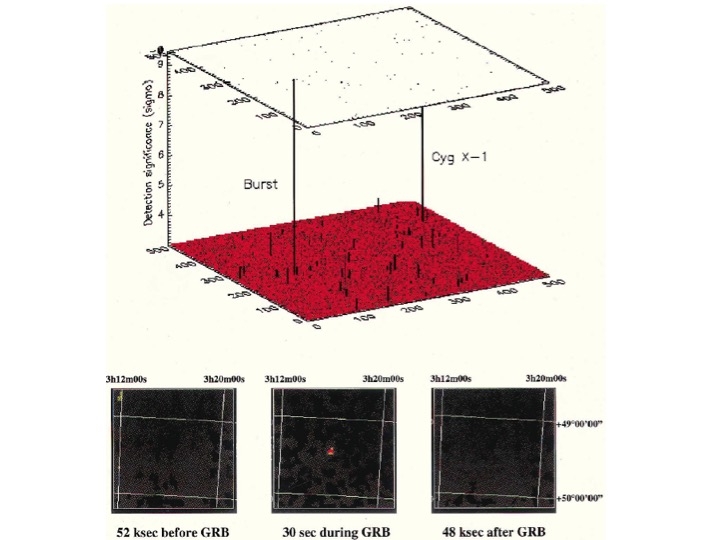}
\caption{The 2-26\ts keV, $40^\circ \times 40^\circ$ discovery images of GRB\ts 960720 obtained by the WFC onboard {\it Beppo-SAX\/}. On top we see the image at the time of the burst integrated over 15\ts s, with Cyg X-1 near the edge of the field; the $y$ axis gives the significance of the detection ($\sigma$). A time sequence of images of the field are shown in the bottom; from \citet{1998A&A...329..906P}}
\label{fig:WFC}
\end{figure}

The Dutch Wide Field Cameras (WFC) \citep{1997A&AS..125..557J} on the Italian mission {\it BeppoSAX\/}  \citep{1997A&AS..122..299B} were responsible for the discovery of the first X-ray afterglow of a gamma-ray burst (GRB\ts 970228) \citep{1997Natur.387..783C}.  The WFCs used a pseudo-random pattern, that has better noise properties than the pure random, and a mask with the same dimensions as the PSDP. In Figure \ref{fig:WFC} we show the coded-mask images taken at the time of GRB960720 and at different times, showing the appearance and disappearance of the burst.

\begin{figure}[!ht]
\centering
\includegraphics[width=0.45\hsize]{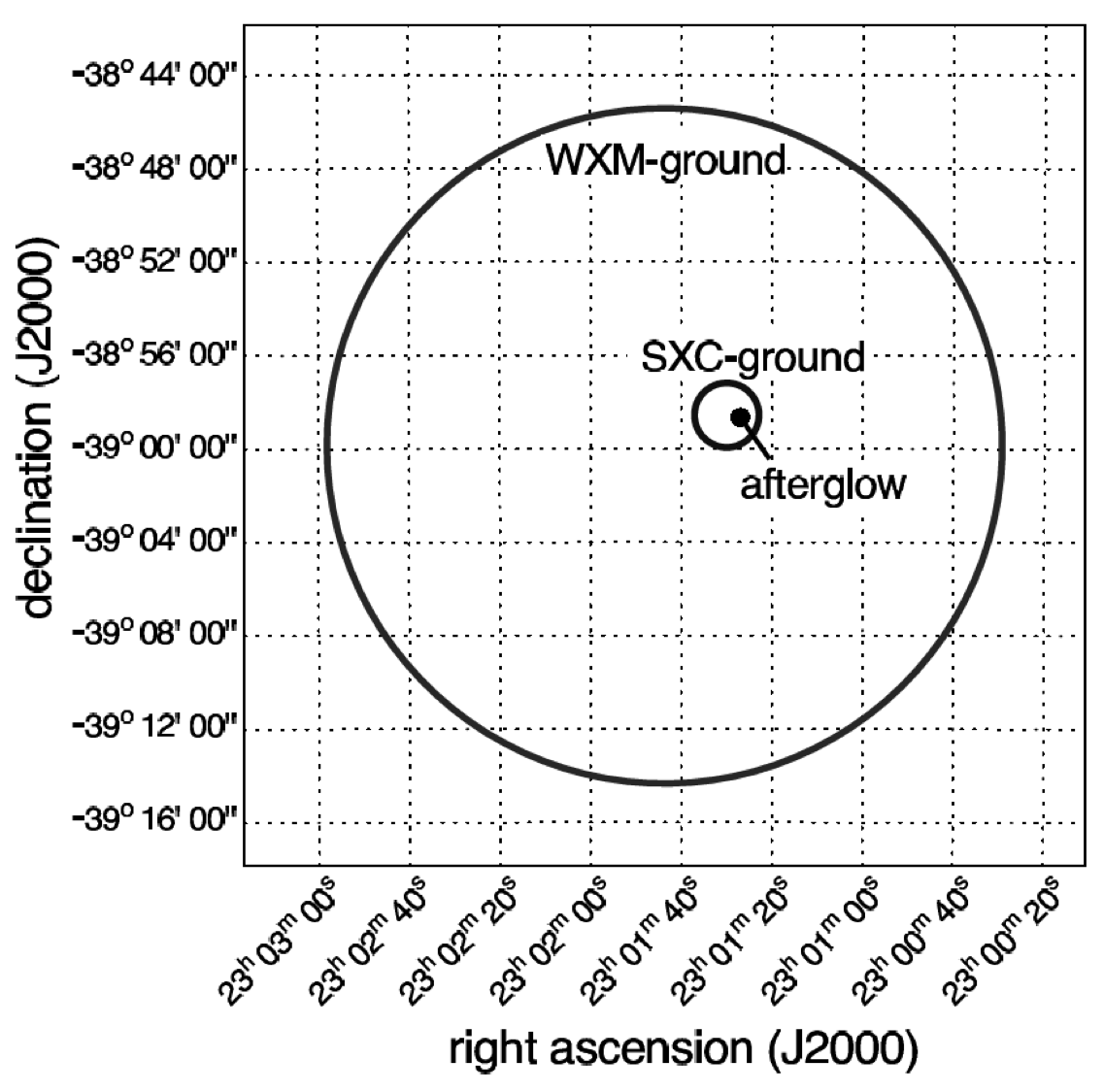}
\caption{A sky map with the {\it HETE\/} localization error circles for GRB\ts 050709 and the location of the X-ray and optical afterglow; from \citet{2005Natur.437..855V}.}
\label{fig:HETE}
\end{figure}

One of the main results of the coded-mask WXM instrument \citep{1995Ap&SS.231..463Y} on the {\it HETE-2\/} satellite \citep{2003AIPC..662....3R} was the observation of the short-hard GRB\ts 050709, whose accurate position determination allowed the identification of the first X-ray and optical afterglows of a short GRB \citep{2005Natur.437..855V}. Figure \ref{fig:HETE} shows a sky map with the {\it HETE-2\/} localization error circles for GRB\ts 050709 and the location of the X-ray and optical afterglow.

\begin{figure}[!ht]
\centering
\includegraphics[width=0.8\hsize]{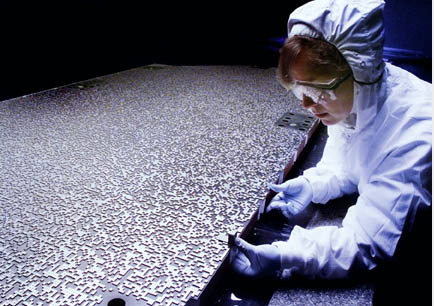}
\caption{The coded mask of BAT/{\it Swift\/} being mounted in the lab; from https://swift.gsfc.nasa.gov.}
\label{fig:BAT_mask}
\end{figure}

\begin{figure}[!ht]
\centering
\includegraphics[width=0.95\hsize]{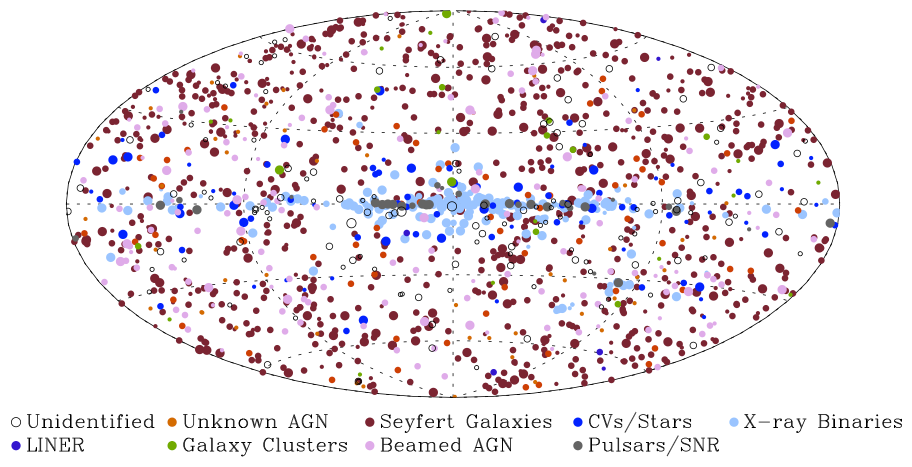}
\caption{A sky map using a Hammer-Aitoff projection in Galactic coordinates showing 1632 sources detected by the BAT coded-mask instrument onboard the Neil Gehrels {\it Swift\/} Observatory. The sizes of the filled dots indicate the measured hard X-ray fluxes; from \citet{2018ApJS..235....4O}.}
\label{fig:BAT_catalog}
\end{figure}

Finally, the BAT instrument \citep{2005SSRv..120..143B} onboard Neil Gehrels {\it Swift\/} Observatory \citep{2004ApJ...611.1005G} has made extremely important discoveries of cosmic explosions in general and GRBs in particular. The instrument has a huge coded mask of 2.7 m$^2$ with 54k lead elements in a random pattern (see Figures \ref{fig:patterns} and \ref{fig:BAT_mask}) and monitors the sky with a FOV of 1.4\thinspace sr and a resolution of 17\ts arcminutes. Figure \ref{fig:BAT_catalog} shows BAT's 105-month hard X-ray survey map \citep{2018ApJS..235....4O} with a great variety of objects detected and positioned by the coded-mask instrument.

\section{Future}
\label{sec:conc}

Coded-aperture imaging has been a powerful technique for acquiring wide-field astronomical images at energies beyond what grazing incidence instruments can achieve. Among other techniques such as Laue lensing \citep{2013IJMPS..23...43V, 2018mgm..conf.3289V} and lobster-eye telescopes \citep{2017SPIE10399E..1QF}, CAI has been one of the preferred choices to study crowded fields of hard X-ray/low-energy $\gamma$-ray sources, and with great success. In particular, coded-mask instruments are very suitable to monitor the transient sky and look for GRBs and other cosmic explosions. Albeit being less sensitive than imaging telescopes in hard X rays by a factor that can reach $\sim \!100$ for a common field (conventional Wolter optics imaging telescopes have in general much smaller fields), CAI is very flexible in the sense that it can be implemented by instruments in different configurations and geometries to achieve very large fields-of-view with wide energy ranges.

Some future X-ray missions such as {\it eXTP} \citep{2019SCPMA..6229502Z} and {\it SVOM\/} \citep{2011CRPhy..12..298P} have coded-mask instruments (WFM and ECLAIR, respectively) to provide wide-field monitoring of the transient sky. These instruments will be able to provide essential coverage of the discovery space so that interesting targets can be found and studied in detail by narrow-field, high-sensitivity focusing instruments. 

Besides CAI, another high-energy wide-field imaging technique that has gained increasing attention is the so-called ``lobster-eye'' telescope. The idea is to mimic the optical system of the eyes of decapod crustaceans, who employ optical reflection (as opposed to refraction in other crustaceans) on a large number of small channels to concentrate light on the sensitive cells, allowing for imaging vision on wide fields-of-view. This technique is very promising in the sense that it overcomes that relative lack of sensitivity of CAI instruments due to its focusing power, whilst retaining the ability to image wide fields.

As an example, the French-Chinese Space-based multi-band astronomical Variable Objects Monitor ({\it SVOM\/}) mission \citep{2011CRPhy..12..298P,2018SPIE10699E..20G}, expected to launch in 2021 to observe Gamma-Ray Bursts, will include the Microchannel X-ray Telescope (MXT) \citep{2018SPIE10699E..21M}. The optics of the MXT is composed of 25 micro-pore optics with plates that have $\sim 600$k square pores of 40\ts $\mu$m size. The inner walls are coated with a 25\ts nm Ir layer to enhance reflectivity. The pnCCD detector effective area at 1\ts keV is 27 cm$^2$. The instrument will operate in the 0.2-10\ts keV range with a FOV of approximately $1^\circ \times 1^\circ$ and an angular resolution $<$ 6 arcmin. The 5$\sigma$ sensitivity is 5 mCrab in 10\ts s and 75 $\mu$Crab in 10 ks. 

Another instrument that will use the lobster-eye technique is the Wide-Field X-ray Telescope on the {\it Einstein probe\/} \citep{2018SPIE10699E..25Y}, a future Chinese mission (to be launched by the end of 2022) that will look for Tidal Disruption Events (TDE) and other transients at soft X-ray energies. The instrument will use micro-pore optics and a BI CMOS detector array with an effective area of 3 cm$^2$. The FOV will be 1.1 sr with an angular resolution of $\sim 5$ arcmin FWHM.

In summary, missions with CAI instruments are still very important in X-ray astronomy due to the unique combination of high observational cadence, very wide fields and large energy bands that it can provide with relatively simple and light instruments. New techniques such as micro-pore (lobster-eye) optics are expected to replace CAI under specific applications for low X-ray energies, with significant gain in sensitivity.

\newpage
\acknowledgements I am grateful to FINEP, CNPq and FAPESP, Brazil, for financial support. I thank Manuel Avila for his contribution to the illustrations. I thank an anonymous referee for very important corrections and suggestions.

 \bibliographystyle{astron} 
 \bibliography{References_paper_review}

\end{document}